# The ground state and polymorphism of $LiSc(BH_4)_4$ finally understood by Density Functional Theory modelling

Agnieszka Starobrat,[a,b] Mariana Derzsi,*[a,c] Tomasz Jaroń,[a] Przemysław Malinowski [a] and Wojciech Grochala[a]

a. Centre of New Technologies, University of Warsaw, Banacha 2c, 02097 Warsaw, Poland.

b. College of Inter-Faculty Individual Studies in Mathematics and Natural Sciences (MISMaP), University of Warsaw, Banacha 2c, 02097 Warsaw, Poland

c. Advanced Technologies Research Institute, Faculty of Materials Science and Technology in Trnava, Slovak University of Technology in Bratislava, 917 24 Trnava, Slovakia.

* corresponding author: mariana.derzsi@stuba.sk

We report a new metastable γ polymorph of mixed metal borohydride $LiSc(BH_4)_4$. Using Density Functional Theory calculations with dispersion corrections, we prove importance of van der Waals H…H interactions for correct theoretical description of the title compound. We propose the ordered ground state structure (α form) and revise the recently reported β phase, now describing it as a solid solution, $LiSc(BH_4)_{4-x}Cl_x$, x≈0.7. The $LiSc(BH_4)_4$ polymorphism is rationalized using $Zr(BH_4)_4$ type structure with Sc → Zr and Li in the interstitial face-centered positions.

Quest for hydrogen-rich compounds has been extremely vivid during the last two decades, with those showing high gravimetric contents of hydrogen being at the focus of the research.[1,2] Borohydrides belong to the family of very H-rich systems with record high gravimetric H content as exemplified by 20.7 wt.% for immensely toxic $Be(BH_4)_2$, followed by 18.4 wt.% for $LiBH_4$, 16.8 wt.% for explosive $Al(BH_4)_3$ and 14.8 wt.% for overly-stable $Mg(BH_4)_2$.[3] To tailor their thermodynamic stability, multi-cation borohydrides have been explored in the recent years, while much attention has been directed towards their synthesis and structural systematics.[4-7] Among them, $LiSc(BH_4)_4$ containing *ca.* 14.4 wt% H represents the first reported alkaline transition-metal bimetallic homoleptic borohydride.[8] Structural characterization of light metal borohydrides is challenging because of low crystallinity, presence of poorly scattering atoms and frequent substitutional disorder. Consequently, until now ground state structure and polymorphism of $LiSc(BH_4)_4$ were not well understood.

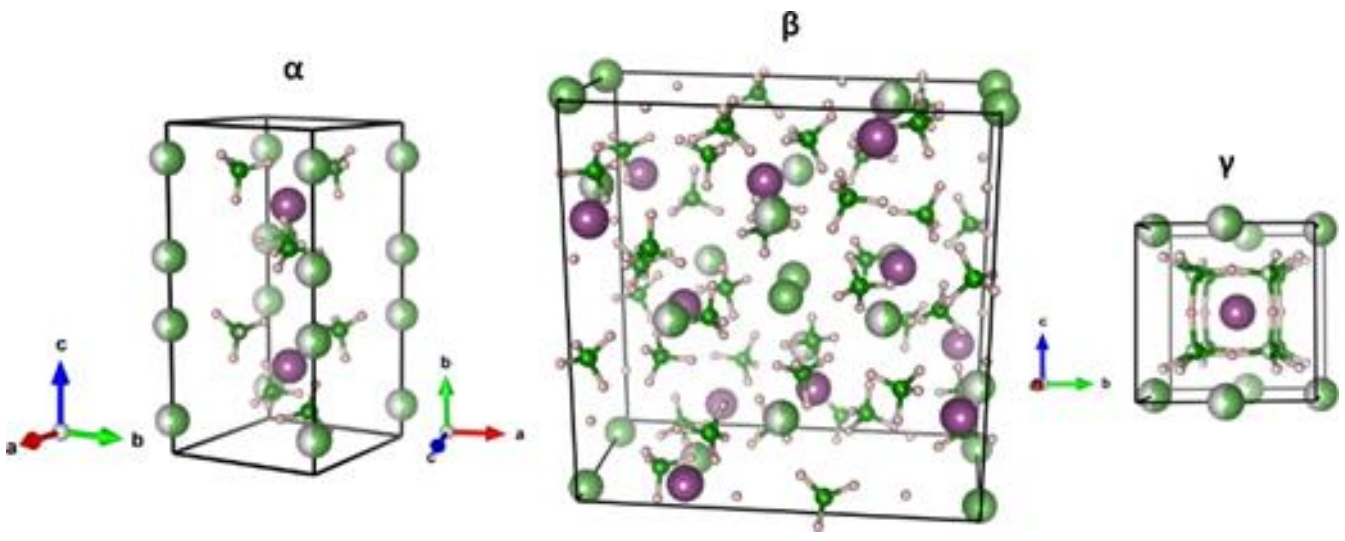

***Figure 1*** *The unit cells of previously reported α- (with Li in 4k positions), β-$LiSc(BH_4)_4$ and new γ polymorph. Colour code: Li – large (half)green, Sc – purple, B – small green, H – light orange balls.*

The first tetragonal α polymorph of $LiSc(BH_4)_4$ with *P*-42c unit cell, V=444.2 Å$^3$ and Z=2 was reported in 2008 (Figure 1).[8] In 2018, second β phase was observed with a tetragonal *I*4/m cell, V= 1504.93 Å$^3$ and Z=8.[9] Here, we describe yet another γ polymorph showing a small cubic *P*-43m cell with V=216.54 Å$^3$ and Z=1[§]. All three phases share the same structural features, namely presence of complex tetrahedral $[Sc(BH_4)_4]^-$ anions and disordered Li sites (Figure 1). Positions of Li (as well as those of H) atoms are difficult to be resolved unambiguously from the X-ray diffraction measurements, and the limited experimental data were insufficient to understand the relative stabilities of the three forms. The problem of Li ordering in the α form was previously addressed employing DFT modelling and phonon direct method.[10] The authors have examined different ordered models with Li atoms either in 2e and 4k Wyckoff positions as suggested earlier[8] and found the 4k positions to be energy-preferred. The corresponding ordered $P222_1$ model, however, does not account for the experimental diffraction pattern. These authors have simultaneously performed a prototype electrostatic ground state search (PEGS) and located a new dynamically stable structure with tetragonal *I*-4 symmetry (distinct from α) and considerably lower energy in respect to the ordered $P222_1$ model (by as much as 404 meV/FU, FU – formula unit). However, such phase has not been observed experimentally. These theoretical findings added to the controversy regarding the crystal structure and stability of the α phase. The subsequent report of the β phase[9] complicated the picture even more. The latter phase spontaneously transforms to α-$LiSc(BH_4)_4$ at T> 120°C, which suggests its metastability with respect to α. However, it is much more densely packed (188.1 $Å^3$ vs. 222.1 $Å^3$), which might suggest lower electronic energy compared to α. Finally, the third polymorphic form, γ, also showing disorder of Li sublattice, has now been discovered (this work[§]). All this calls for theoretical re-examination of the polymorphism of $LiSc(BH_4)_4$. Here, we finally settle the dispute using dispersion-corrected Density Functional Theory (DFT-D3) modelling and highlight the crucial importance of van der Waals (D3) correction to reach agreement with experiment[‡].

One key structural feature of α and γ form is simple cubic (or slightly distorted) packing of the complex $[Sc(BH_4)_4]^-$ anions, which is typical also of homoleptic borohydrides $M(BH_4)_4$ (M = $Zr^{4+}$ or $Hf^{4+}$).[11] In $LiSc(BH_4)_4$ polymorphs, the cubic symmetry is bent mostly by the presence of Li counterions in interstitials. We therefore use □$Zr(BH_4)_4$ prototype (where □ stands for interstitial site) as a starting point for modelling of crystal structures of the $LiSc(BH_4)_4$ polymorphs by Zr → Sc substitution and allocation of $Li^+$ in diverse interstitial positions (Figure 2).

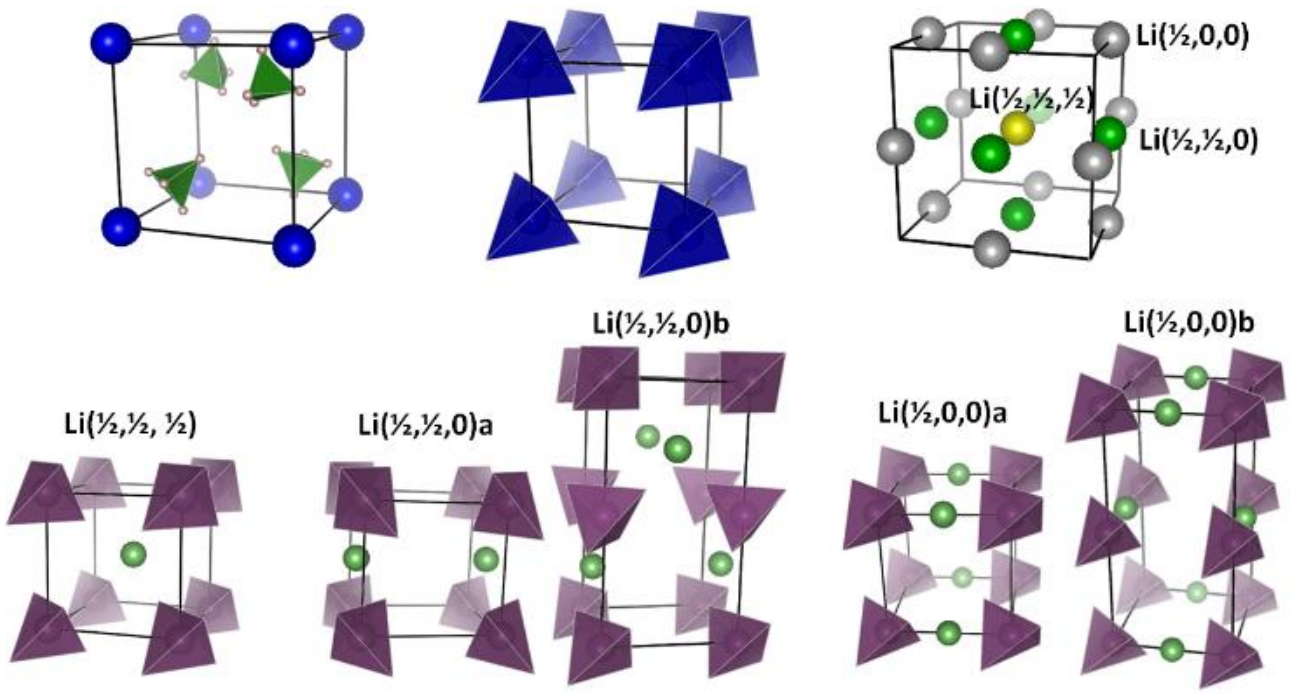

***Figure 2*** *TOP – Cubic P-43m $Zr(BH_4)_4$ unit cell (top left) highlighting the $[ZrB_4]$ tetrahedra (top middle) and tetrahedral interstitial sites, □, available for Li (top right). BOTTOM - hypothetical $LiSc(BH_4)_4$ models in □$Zr(BH_4)_4$ type structure (Sc→Zr, Li → □). Each model represents filling of different interstitial by Li. Colour code: blue/purple tetrahedra ((Zr/Sc)$B_4$), green tetrahedra ($BH_4$), bluegreen/purple balls (Zr/Li/Sc). See text for details.*

The cell offers three non-equivalent interstitial tetrahedral sites for the lithium counter cation at the edge (½,0,0), face (½,½,0) and centre (½,½,½), with tetrahedral coordination by four borohydride groups on each site (Figure 2). The centre (½,½,½) position accounts for CsCl-type ordering of the Li-Sc sublattice that was demonstrated to account for majority of the strong XRD reflections of α phase.[8]

Furthermore, this model (represented by the *P*-43m structure in Figure 2) seems to account very well for all XRD reflections of the new γ phase. Therefore, we have examined dynamical stability of the *P*-43m structure by computing its phonon dispersion curves and searched for possible lower-energy phonon-induced distortions associated with the imaginary (destabilizing) phonons (see S1 in ESI).[12,13]

Altogether, we have systematically examined seven models derived from $Zr(BH_4)_4$ structure. They account for one- to -three dimensional networks with CsCl, ZnO, CuO and NaCl type Li-Sc sublattices. Their key structural features including coordination polyhedra and type of connectivity are listed in Table 1.

The most important findings of the computations are as follows:

(i) The *P*-42c model originating from face-centred site occupation in Li(½,½,0)b (Figure 3) was found to have the lowest energy of all examined models, lower even than the previously suggested ground state *I*-4 polytype.[10] The high-symmetry *P*-43m model is unstable with respect to our best structure by 655(701) meV/FU (DFT/DFT-D3). Importantly, our *P*-42c ground state model yields better Rietveld fit for the α form[§§§] than the literature *P*-42c model (cf. Fig. S3 and S4 in ESI);

(ii) The *I*-4 structure predicted previously with PEGS method and the $P222_1$ structure considered preciously as the ordered model of α,[10] were obtained also in this study by following the imaginary modes in the *P*-43m model (cf. S1 in ESI).

(iii) Our quest yielded tens of structures obtained by following the imaginary modes in the *P*-43m model (not shown) but none could explain the diffraction pattern of the γ phase. Therefore, the γ phase, which yields nearly identical XRD pattern as the α one (cf. Fig. S8 in ESI), seems to exhibit intrinsic substitutional disorder[§§§], which serves as a stabilizing factor for this phase.

(iv) Despite extensive quest towards the β phase, no structure model was found, which would satisfactorily describe the lattice parameters and volume of this phase (for possible ordered models see Fig S5 in ESI).

***Table 1*** *Selected structural features of the computed $LiSc(HB_4)_4$ models derived from □$Zr(BH_4)_4$ prototype. In all models, $[LiB_4]$ tetrahedra are present except the $P222_1$ structure where Li is in kinked $[LiB_2]$ coordination. Sqr = square, tetra = tetrahedral. Labelling of the models is explained in Fig. 1; X,Γ stand for structures obtained following the dynamically unstable modes in respective points in the Brillouin zone. See text for further details.*

| | | | Coordination polyhedra | | |
|---|---|---|---|---|---|
| **model** | **LiSc sublattice** | **H-Li$_x$** | **[LiH$_x$]** | **[LiSc$_x$]** | **[ScLi$_x$]** |
| Li(½,½,½) | CsCl | 1-dentate | tetra | cubic | cubic |
| Li(½,½,½)X | deformed CsCl | 1-,2-dent. | 6-fold | 4-fold | sqr |
| Li(½,½,½)Γ | ZnS | 2-dentate | 8-fold | tetra | tetra |
| Li(½,0,0)a | 1D – parallel chains | 2-dentate | 8-fold | linear | linear |
| Li(½,0,0)b | 2D – perpendicular chains | 2-dentate | 8-fold | linear | Linear |
| Li(½,½,0)a | 2D-NaCl | 2-dentate | Sqr prism | sqr | sqr |
| Li(½,½,0)b | CuO | 2-dentate | Sqr prism | sqr | tetra |

It turns out that the interionic H…H contacts have important impact on the crystal structure and stability of all studied $LiSc(BH_4)_4$ models and their proper treatment within the DFT-D3 framework[‡] was found to be indispensable in order to reach qualitative agreement between the predicted *P*-42c ground state and diffraction data for α phase (see S3 in ESI). To demonstrate this fact, in Table 2 we compare the DFT and DFT-D3 relative energies, lattice parameters the shortest interionic H…H contacts fo r all models. The DFT-D3 correction for the weak dispersive interactions has led to considerable volume reduction (6–20%) as compared to standard

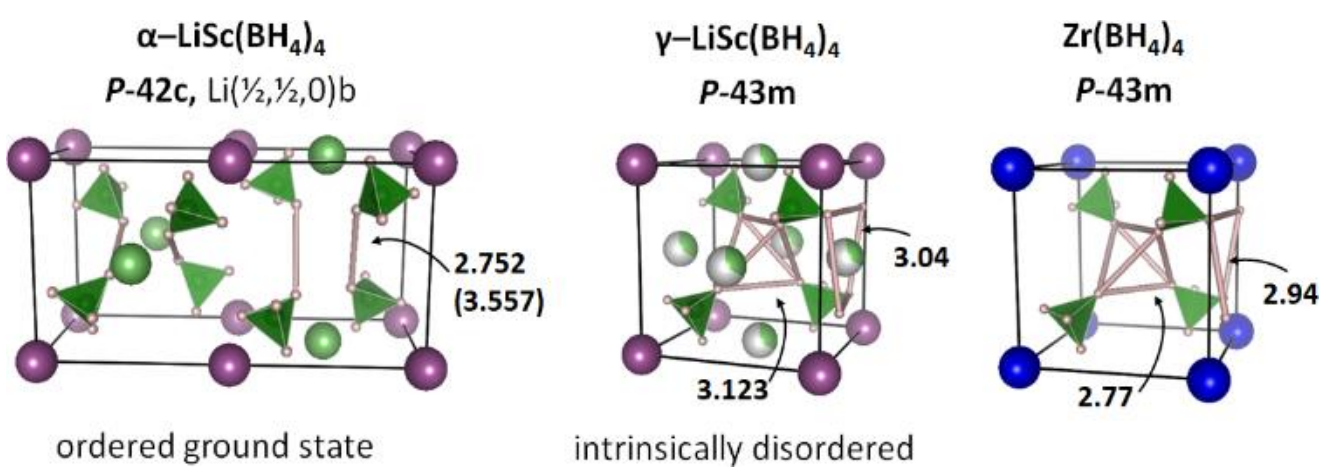

***Figure 3*** *Predicted ordered model of α–LiSc(BH$_4$)$_4$, new intrinsically disordered γ–LiSc(BH$_4$)$_4$ and Zr(BH$_4$)$_4$ unit cell highlighting selected secondary intermolecular H...H contacts (light orange bonds). Experimental values are provided for γ and the Zr(BH$_4$)$_4$ structure[14] and DFT-D3 (DFT) for α. All structures are shown in Zr(BH$_4$)$_4$ representation. Colour code: BH$_4$ - green tetrahedra, Zr - blue, Li - green, Sc - purple, H - light orange balls.*

DFT approach. This is mostly due to substantial shortening of the intermolecular H…H contacts across the empty voids* from *ca.* 3 Å and longer (DFT) to 2.425–2.838 Å (DFT-D3). The shortest DFT-D3 values are comparable to twice the van der Waals radius of H (2.4 Å), while the longer ones are within the range observed in the Zr(BH$_4$)$_4$ and Hf(BH$_4$)$_4$ crystals (2.77–2.94 Å).[14-16] Note, that in case of the ground state *P*-42c structure, model Li(½,½,0)b, plain DFT greatly underestimates the H…H interactions (3.557 Å) relative to DFT-D3 (2.752 Å) and the volume reduction due to dispersive interactions amounts to 17%. The network of selected H…H contacts for the ordered model of α, disordered γ and Zr(BH$_4$)$_4$ is shown in Figure 3.

Impact of the intermolecular H…H interactions on the stability of the structures is manifested also by the comparison of their DFT and DFT-D3 energies calculated relative to the parent *P*-43m structure with Li(½,½,½) occupancy (Table 2). For the ground state *P*-42c structure with Li(½,½,0)b occupation, the relative energy lowers by additional 46 meV with inclusion of the D3 correction. Comparable energy lowering is obtained also for the *I*-4 structure (42 meV), while in case of the remaining models the energy lowering is even larger (61–174 meV), consistently with changes in the H…H separations (further details in Figure S9 in ESI).

As already mentioned, the unusual low-volume β form could not be reproduced by our DFT-D3

***Table 2*** *List of cell parameters (a, b, c in Å; V/Z in Å$^3$), shortest intermolecular H...H separations d(H...H)$_{min}$ (Å) and relative energies E (meV/FU) of the LiSc(HB$_4$)$_4$ models derived from □Zr(BH4)4 prototype calculated with (DFT-D3) and without (DFT) van der Waals correction at zero (p,T) conditions and respective experimental data for α and γ polymorph. The labelling of the models is the same as in Table 1. The experimental values for the α phase are obtained by fitting its XRD pattern with the P-42c ground state optimized with DFT-D3 (the DFT optimized model was not). For further details see the text. SPGR = space group, Z = number of formula units in the unit cell, V/Z = volume per one formula unit.*

| | | | DFT | | | | | | DFT-D3 | | | | | |
|---|---|---|---|---|---|---|---|---|---|---|---|---|---|---|
| **model** | **SPGR** | **Z** | **V/Z** | **a** | **b** | **c** | **d(H…H)$_{min}$** | **E** | **V/Z** | **a** | **b** | **c** | **d(H…H)$_{min}$** | **E** |
| Li(½,½,½) | *P*-43m | 1 | 233.7 | 6.159 | 6.159 | 6.159 | 2.974 | 0 | 218.7 | 6.025 | 6.025 | 6.025 | 2.832 | 0 |
| Li(½,½,½)X | *P*222$_1$ | 2 | 223.6 | 6.067 | 6.072 | 12.137 | 2.646 | -215 | 193.8 | 5.793 | 5.714 | 11.711 | 2.425 | -350 |
| Li(½,½,½)Γ | *I*-4 | 2 | 248.3 | 6.410 | 6.410 | 12.050 | 3.338 | -638 | 198.8 | 5.760 | 5.760 | 11.985 | 2.800 | -680 |
| Li(½,0,0)a | *P*-42m | 1 | 243.5 | 6.272 | 6.272 | 6.189 | 2.996<br>3.262 | -194 | 193.6 | 5.651 | 5.651 | 6.060 | 2.425<br>2.479 | -275 |
| Li(½,0,0)b | *P*-42c | 2 | 233.7 | 6.206 | 6.206 | 12.528 | 2.959 | -193 | 198.8 | 5.959 | 5.959 | 11.194 | 2.503 | -254 |
| Li(½,½,0)a | *P*-42m | 1 | 225.0 | 5.971 | 5.971 | 6.308 | 3.124 | -434 | 188.0 | 5.837 | 5.837 | 5.518 | 2.449 | -608 |
| Li(½,½,0)b ground state | *P*-42c | 2 | 243.2 | 6.374 | 6.374 | 11.973 | 3.557 | -655 | 202.5 | 5.851 | 5.851 | 11.827 | 2.752 | -701 |
| | | | | | *Experimental data* | | | | | | | | | |
| α ordered$^{T=300K}$ | *P*-42c | 2 | 221.2 | 6.067 | 6.067 | 12.015 | 3.02 | | | | | | | |
| γ disordered$^{T=100K}$ | *P*-43m | 1 | 216.5 | 6.005 | 6.005 | 6.005 | 3.04 | | | | | | | |

calculations. This failure, as well as the fact that the volume of what was believed to be a β form is much smaller than those of the α or γ ones, motivated us to re-investigate the β phase. We have now allowed for partial incorporation of the smaller $Cl^-$ anions into the positions of larger $BH_4$ moieties[§§]. Such approach results in an improved Rietveld fit in comparison to the model assuming purely borohydride-based system (Fig. S7 in ESI). The revised chemical formula of the crystalline phase previously assigned as β-$LiSc(BH_4)_4$[9] is in fact $LiSc(BH_4)_{4-x}Cl_x$, where x≈0.7[§§§]. Thus, chloride contamination in this mixed-anion $BH_4$-Cl phase is substantial as is the case for a number of borohydride-halide systems.[17-22]

## Conclusions

We have prepared a new metastable γ polymorph of $LiSc(BH_4)_4$, which added to two previously reported forms, α and β. We have been able to resolve the old standing problem of the structure and stability of the α polymorph of this compound using DFT-D3 modelling. Its structure is best described by the ordered *P*-42c model, which has also the lowest computed energy among all forms studied and thus confirms the ground state character of the α polymorph. The new γ form, which has been characterized here using a single crystal diffraction, shows a substantial substitutional disorder, which is a stabilizing factor for this structure. The XRD pattern for the γ form is similar to that of the ordered α one, which agrees with similar heavy atom sublattices (Li, Sc) of both forms.

Our systematic study utilizing the cubic □$Zr(BH_4)_4$ structure models with Sc → Zr substation and Li filing the voids, has revealed the ground state *P*-42c structure and importance of the H…H contacts on crystal structure and stability of quasi-molecular borohydrides containing light metal cations and necessity to account for them in DFT calculations. The chemical identity of the previously reported β phase was put into question based on the DFT and DFT-D3 calculations, and consequently redetermined as mixed anion $BH_4$-Cl phase, $LiSc(BH_4)_{4-x}Cl_x$, where x≈0.7[§§].

## Acknowledgements

This research was carried out with the support of the Interdisciplinary Centre for Mathematical and Computational Modelling, University of Warsaw under grant no. ADVANCE++ (GA76-19). M.D. acknowledges the European Regional Development Fund, Research and Innovation Operational Programme, for project No. ITMS2014+: 313011W085. The authors acknowledge the support from Polish National Science Centre under grant HYDRA no. 2014/15/B/ST5/05012. The authors thank the Biopolymers Laboratory, Faculty of Physics, University of Warsaw, for the access to Agilent Supernova X-ray single-crystal diffractometer, co-financed by the European Union within the ERDF Project POIG.02.01.00-14-122/09.

## Notes and references

‡ Density Functional Theory (DFT) calculations utilizing PBE functional[23] were performed with the projected-augmented-wave (PAW) method, as implemented in VASP 5.4 code.[24-27] Valence electrons (Li: 1s2s2p, Sc: 3p4s3d, B: s2p1 and H: ultrasoft test) were treated explicitly, while standard VASP pseudopotentials (accounting for scalar relativistic effects) were used for the description of core electrons. The cut-off energy of the plane waves was set to 650 eV, a self-consistent-field convergence criterion to $10^{-7}$ eV (electronic) and $10^{-5}$ eV (ionic cycle), Gaussian smearing width to 0.05 eV, and the k-point mesh was set to 0.25 $Å^{-1}$. Correction for van der Waals interactions was treated by DFT-D3 method with Becke-Jonson damping.[28] We have tested this method on $Zr(BH_4)_4$ crystal and it provided excellent agreement with experimental volume (within 1%) while simple DFT overestimated it by 9%.

Lattice dynamics (phonons) was calculated using direct method implemented in the program PHONOPY.[29] The input Hellman-Feynman forces were calculated for 2x2x2 supercell with PBE functional in VASP program.

§ The γ polymorph single crystals were obtained in the reaction $ScCl_3$ + $3LiBH_4$ in the environment of solvent (DMS) during its slow evaporation. Fast evaporation led to polycrystalline α polymorph. Detailed information on synthesis method can be found in the study on similar $MSc(BH_4)_4$ systems.[30] The crystals were measured at 100K on Agilent Supernova X-ray diffratometer with Cu-Kα radiation (microsource). Crystals were covered in Krytox 1531 perfluoro polyalkyl ether oil. Data collection and reduction was performed with CrysAlisPro software (v. 38.43).[31] Structure solution: SHELXT,[32] refinement against $F^2$ in Shelxl-2013,[33] with ShelXle as GUI software.[34]

§§ Structure was re-refined using Jana2006 software[35] using β form as starting model with all restrains as described before.[9] Cl atoms were put additionally into B positions with sum of B+Cl occupancies equal to 1 and hydrogen occupancies depending on B ones. Occupancies for independent atomic positions were kept identical. The Rietveld fit and differential profile are shown in the ESI together with numerical parameters of the fit.

§§§ The details of the crystal structure of ordered model of α, disordered γ and revised β phase may be obtained from the CCDC Database (https://www.ccdc.cam.ac.uk/structures/) on quoting the deposition numbers 2007629 for α-$Li[Sc(BH_4)_4]$, 1890079 for β-$Li[Sc(BH_4)_{3.31}Cl_{0.69}]$, and 2007744 for γ-$Li[Sc(BH_4)_4]$. For pre-publication data contact structures@ccdc.cam.ac.uk

* The discussion refers exclusively to the crystallographic directions, where other type of contacts is absent (see S9 in ESI).

# Supplementary Information

S1. Graphical algorithm of obtaining all structures

S2. DFT calculated phonon dispersion curves of the Li(½,½,½) model

S3. Simulated diffraction patterns of α: DFT and DFT-D3 performance

S4. Rietveld fits of the diffraction patterns of α phase (disordered and ordered model)

S5. Calculated ordered models of the beta phase

S6. Redetermination of the beta phase: Rietveld fits of the diffraction patterns without and with partial Cl → $BH_4$ substitution

S7. The single crystal data for the disordered γ phase and its simulated powder diffraction pattern

S8. CIF of theoretically predicted ordered model of α

S9. The secondary intermolecular H…H contacts in selected calculated models

S10. References

**S1. Graphical algorithm of obtaining all structures.**

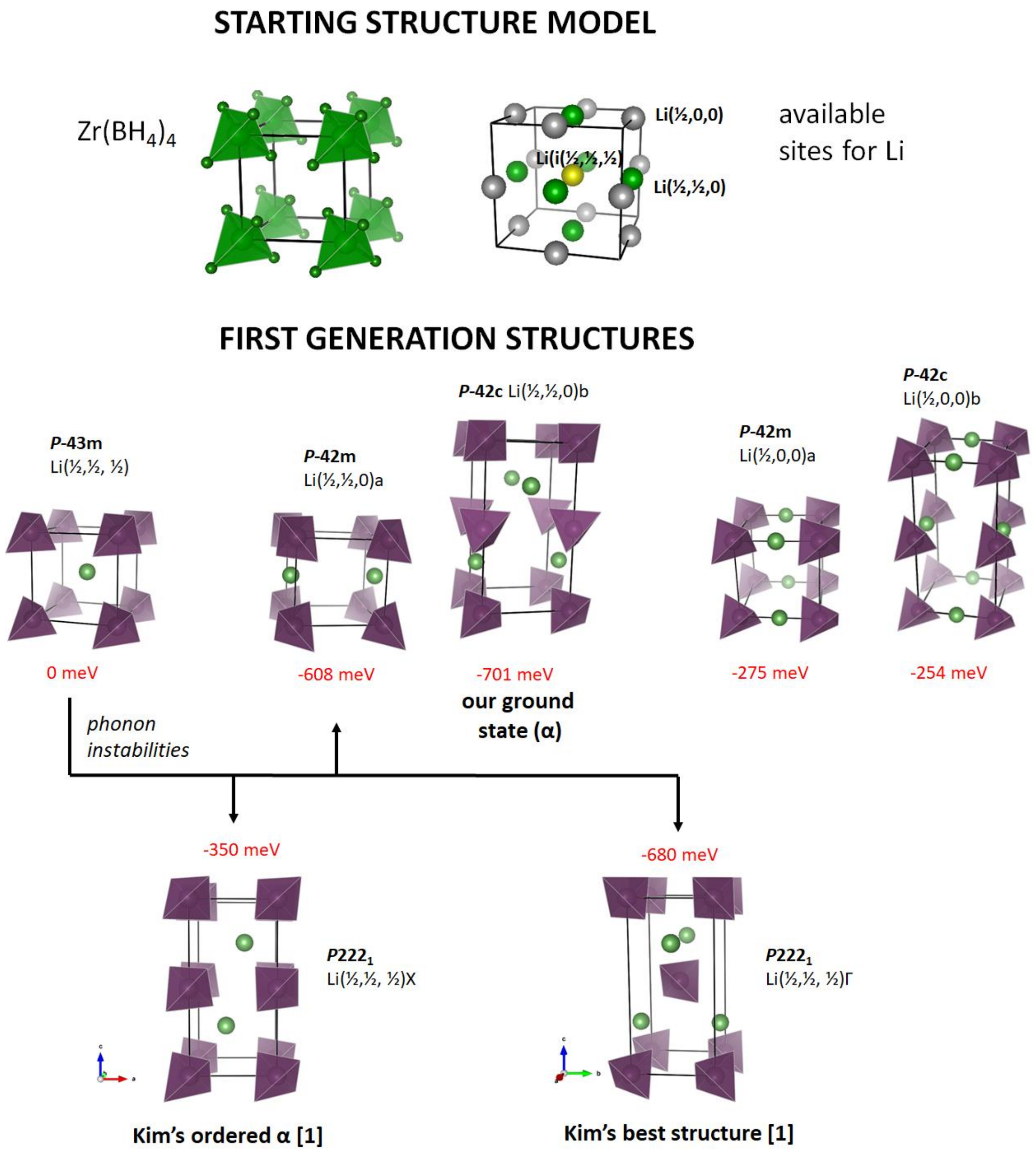

## S2. DFT calculated phonon dispersion curves of the Li(½,½,½) model

We have calculated phonon dispersion curves for $LiSc(BH_4)_4$ in $Zr(BH_4)_4$ type structure with Li placed in the (½, ½, ½) position (Figure S1). This model is dynamically unstable as manifested by four optical modes that gain negative energies across the entire Brillouin zone. We have searched for lower energy solutions by following the distortions along all special points: Γ(0,0,0), M(½, ½,0), X(0, ½,0) and R(½, ½,½). Namely, we have distorted the original structure along the modes and ran full DFT optimization. Tens of structures were obtained but none could explain the diffraction data of the α, β or γ phase. Importantly, all these structures have higher energies in respect to the *P*-42c model derived from the □$Zr(BH_4)_4$ prototype with Li(½,½,0)b occupancy that represents the ground states structure found in this study.

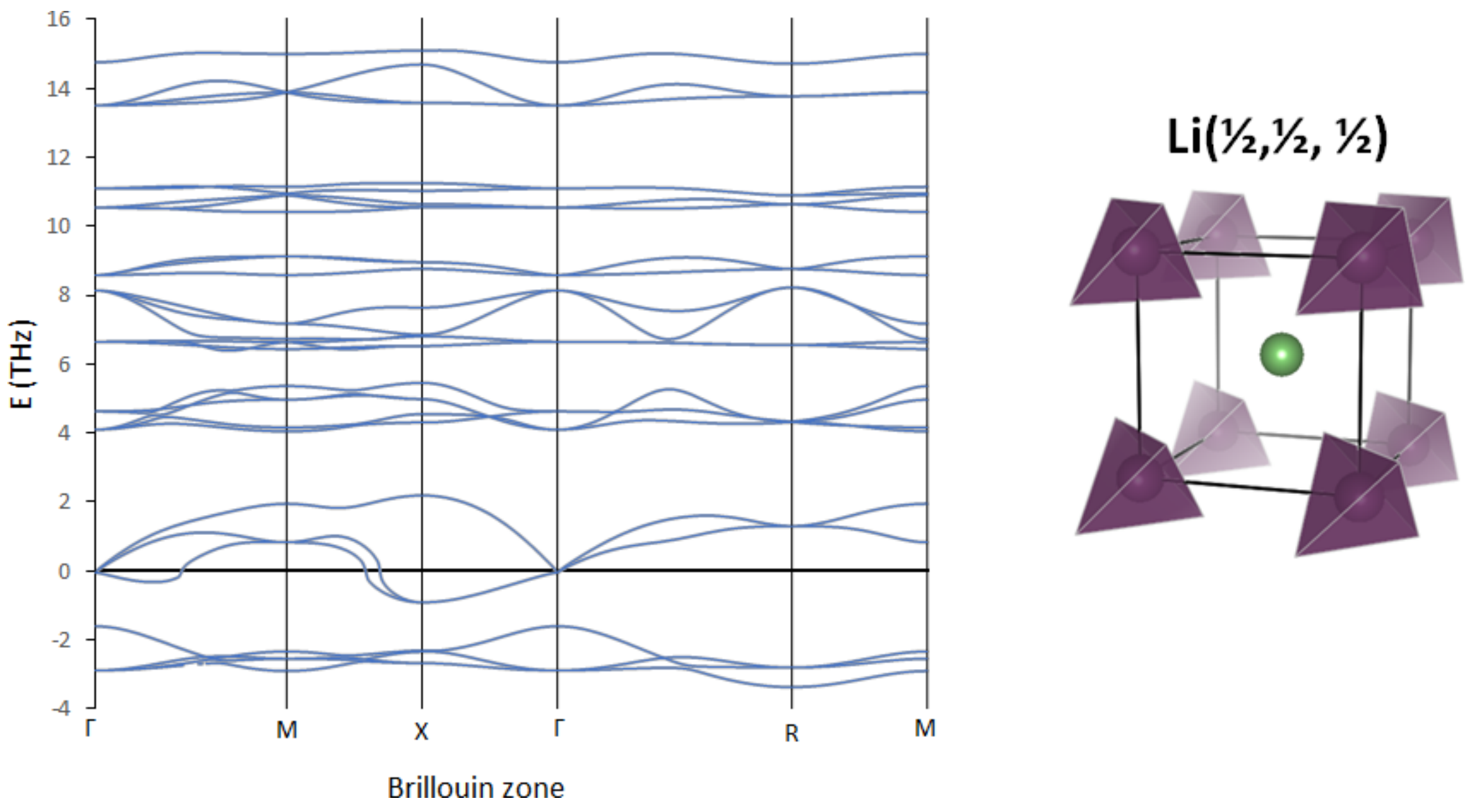

**Figure S1:** Phonon dispersion curves (left) calculated for $Zr(BH_4)_4$ type model of $LiSc(BH_4)_4$ with Li in the (½, ½, ½) position (right). Color code: $Li^+$ – green ball, complex $[Sc(BH_4)_4]^-$ anions – purple tetrahedra. The vortexes of the tetrahedra represents $BH_4$ groups.

## S3. Simulated diffraction patterns of α: DFT and DFT-D3 performance

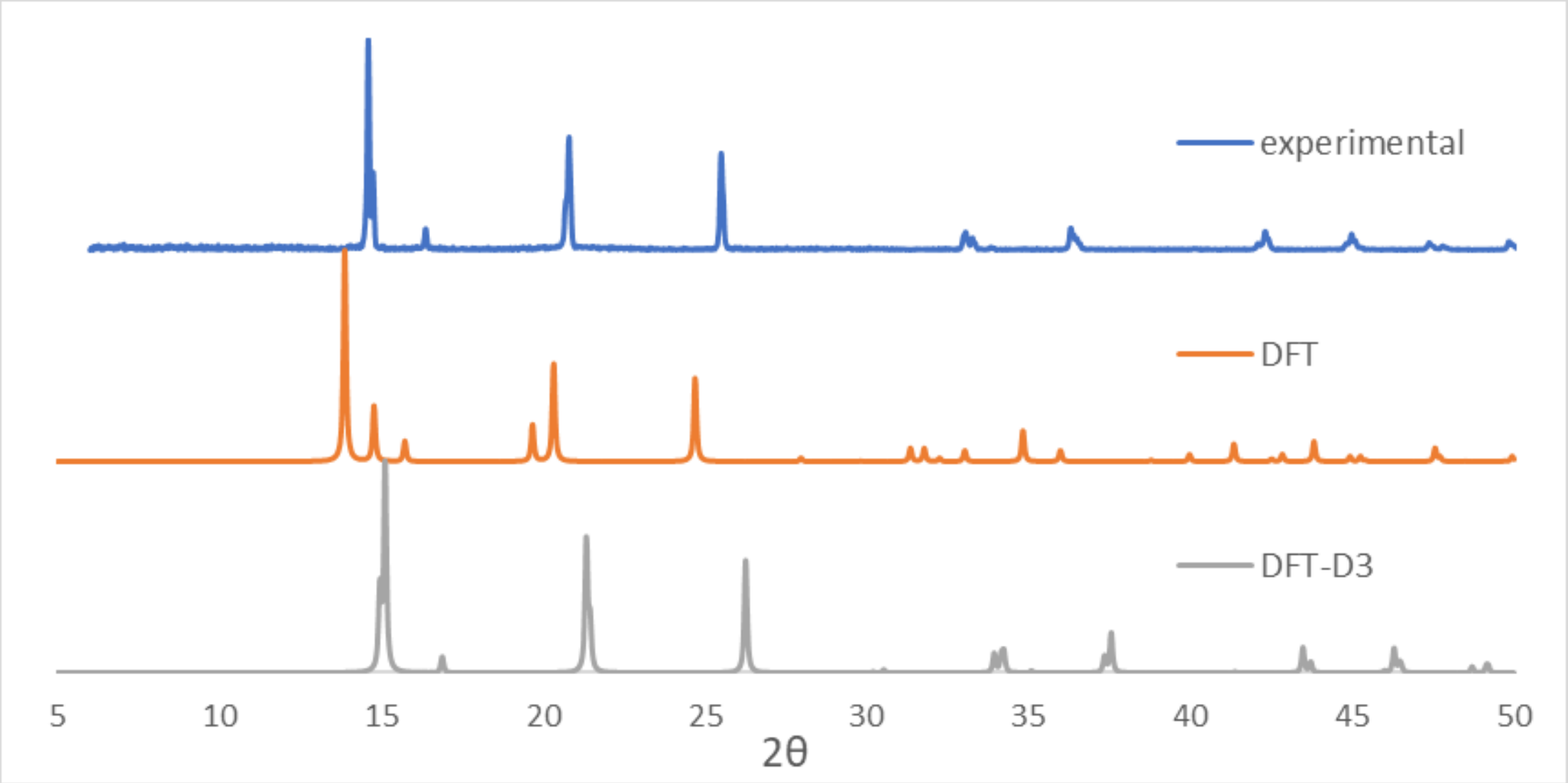

**Figure S2:** Comparison of the experimental diffraction pattern of α phase with simulated diffraction pattern of ground state model *P*-42c, as calculated at DFT and DFT-D3 level.

## S4. Rietveld fits of the diffraction patterns of α phase (disordered and ordered model)

As simulated diffraction pattern of *P*-42c ordered ground state seems to successfully reproduce experimental pattern of α phase, Rietveld refinement of a polycrystalline sample containing this phase was re-examined. Results of two refinements were compared showing that fit parameter wRp gets lower while using ordered model than while using previously reported disordered model of α phase.

During both refinements, several restrains concerning H atoms positions and atomic displacement parameters (ADP) were defined. $BH_4$ groups were fixed in a tetrahedral geometry with B-H distances set as 1.15 Å (tolerance of 0.01 Å) and H-B-H angles restrained to 109.47° (tolerance of 0.01°). ADP of all H atoms are equal and 1.2 times larger than ADP of B atom.

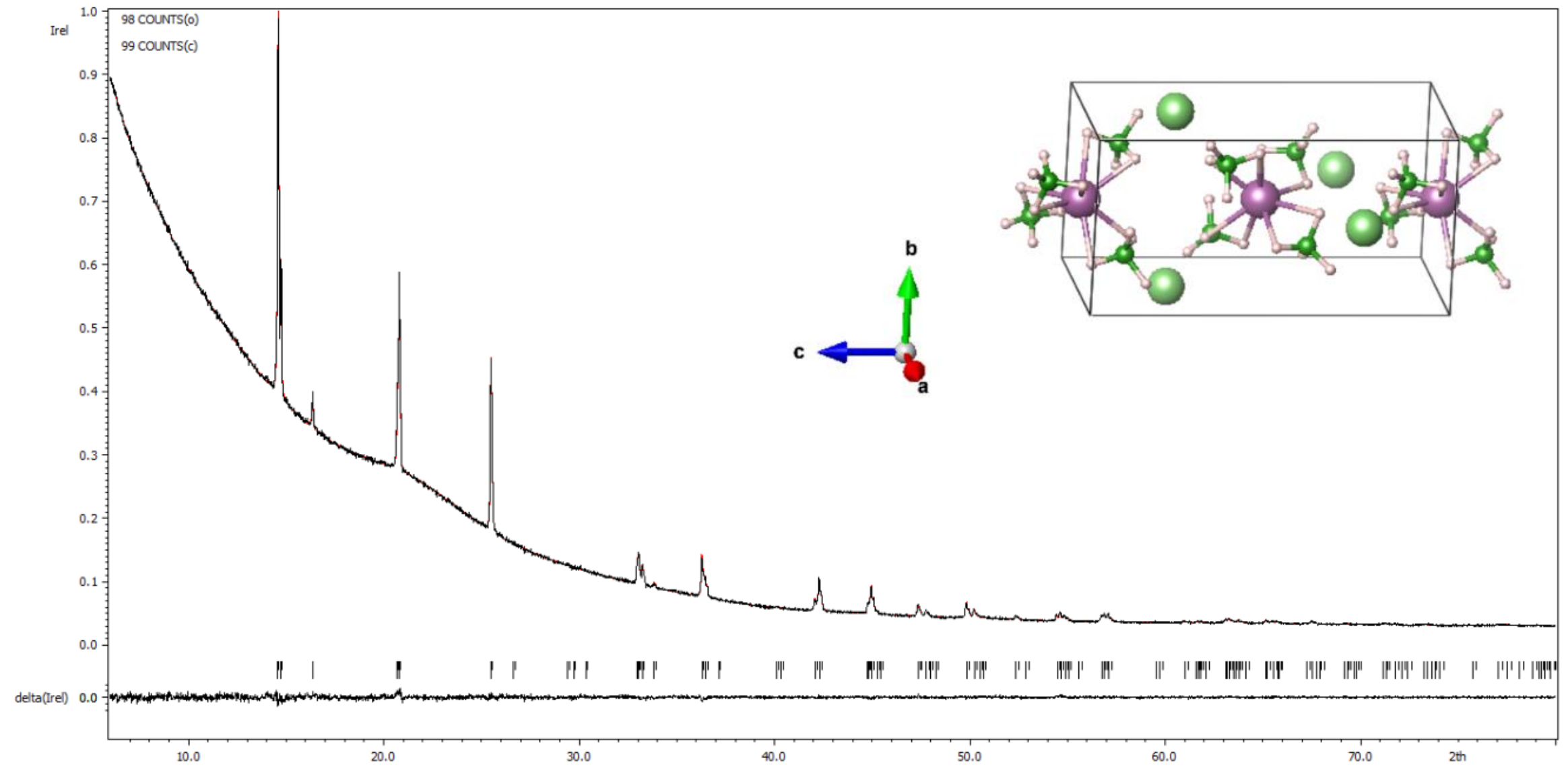

**Figure S3.** The results of Rietveld refinement of α-$LiSc(BH_4)_4$ using ordered *P*-42c model (the ground state found in this study). Inset: resulting structure, color code: Sc – purple, Li – big green, B – small green, H – light orange.

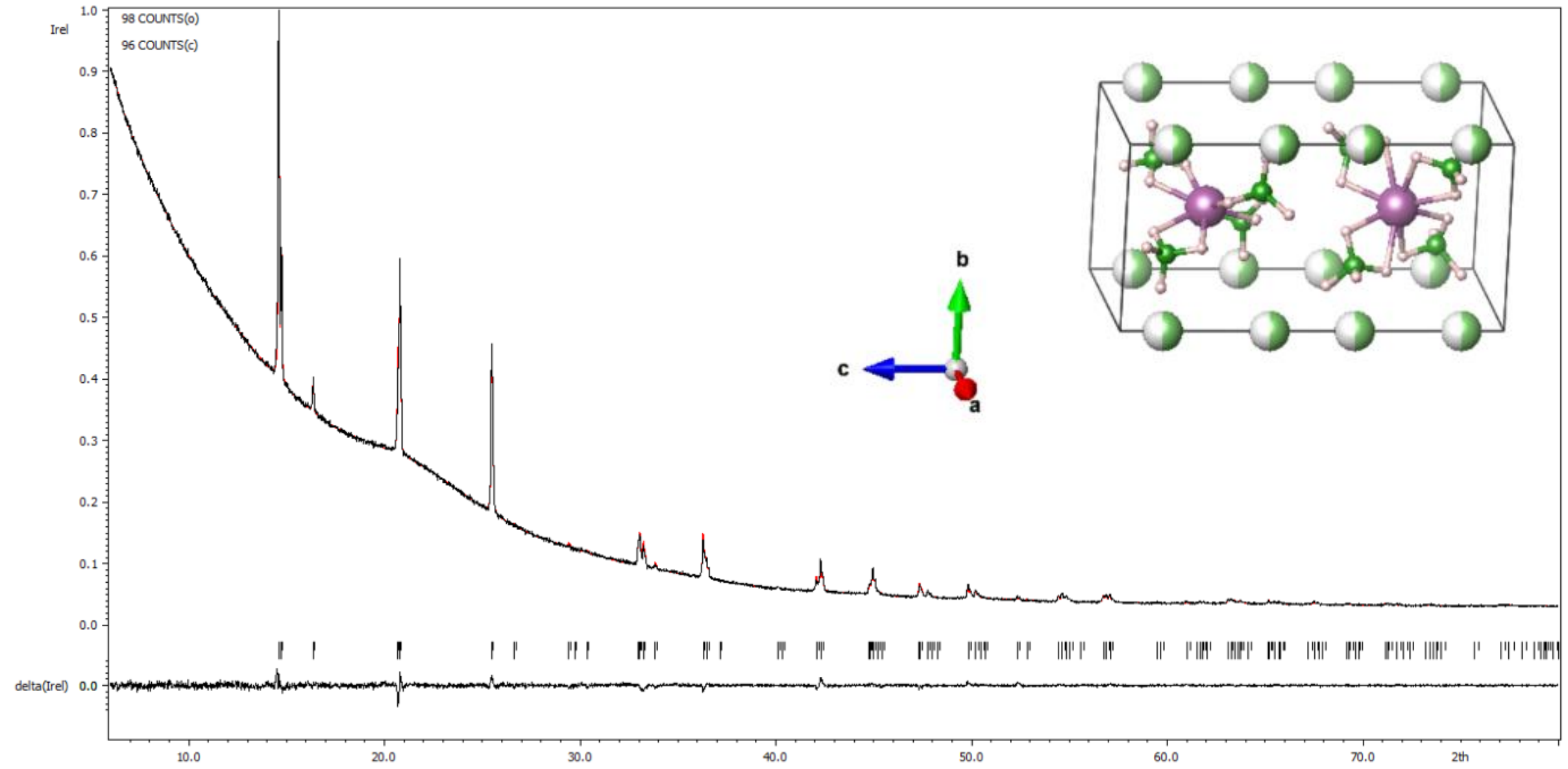

**Figure S4.** The results of Rietveld refinement of α-$LiSc(BH_4)_4$ using disordered *P*-42c model [2]. Inset: resulting structure, color code: Sc – purple, Li – big half-green, B – small green, H – light orange.

**Table S1.** Comparison of refined cell parameters and wRp/cRp of fits obtained using ordered and disordered models of α phase.

| | **ordered model** | **disordered model** |
|---|---|---|
| **SPGR** | *P*-42c | *P*-42c |
| **wRp [%]** | 1.17 | 1.44 |
| **cRp [%]** | 28.63 | 32.21 |
| **a [Å]** | 6.0670(5) | 6.0710(8) |
| **c [Å]** | 12.0147(10) | 12.0233(16) |

## S5. Calculated ordered models of the beta phase

Crystal structure of the β phase was originally resolved in tetragonal *I*4/m space group with half occupancies assigned to half of the lithium atoms. To validate the structure, we have built and optimized several ordered models following two approaches. In the first one, we have constructed two ordered models based on originally determined I4/m structure by removing half of the partially occupied lithium positions in the unit cell (model **1**) and supercell 1x1x2 (model **2**) respectively. The experimentally refined unicell contains eight lithium positions with half occupancy. These positions are numbered in Figure S5. In model **1** we have removed Li positions 1 to 4 (or alternatively 5-8). This model preserves the original *P*4/m symmetry of β phase. Model **2** was built by removing the following Li atoms from a 1x1x2 supercell: atoms 2 and 4 with z=0, atoms 5 and 7 with z=0.25, atoms 1 and 3 with z=0.5 and 6 and 8 with z=0.75. This model has space group $P4_2/n$.

Second approach is based on our observation that β is a thermal decomposition product of $NH_4Sc(BH4)_4$. Here, we have assumed that $NH_4^+$ evolved from the sample while being heated and LiCl manifested its presence by substituting ammonium cation with Li. The $NH_4Sc(BH4)_4$ type structure is partially supported by the fact that a close-to-tetragonal β type representation of the $NH_4Sc(BH4)_4$ crystal exists (compare first two columns in Table S2). Model **3** was constructed by $NH_4$ → Li substitution. Models **4** was built by transforming the $NH_4Sc(BH4)_4$ structure to β representation using transformation matrix (2 0 0) (0 -1 1) (0 1 1)and subsequently removing half of the formula units (one $NH_4Sc(BH4)_4$ layer).

All models are illustrated in Figure S5. Energies and lattice parameters of all models of β $LiSc(BH_4)_4$ are compared in Table S2. None of these modes satisfy the observed XRD pattern of the beta phase (Figure S6).

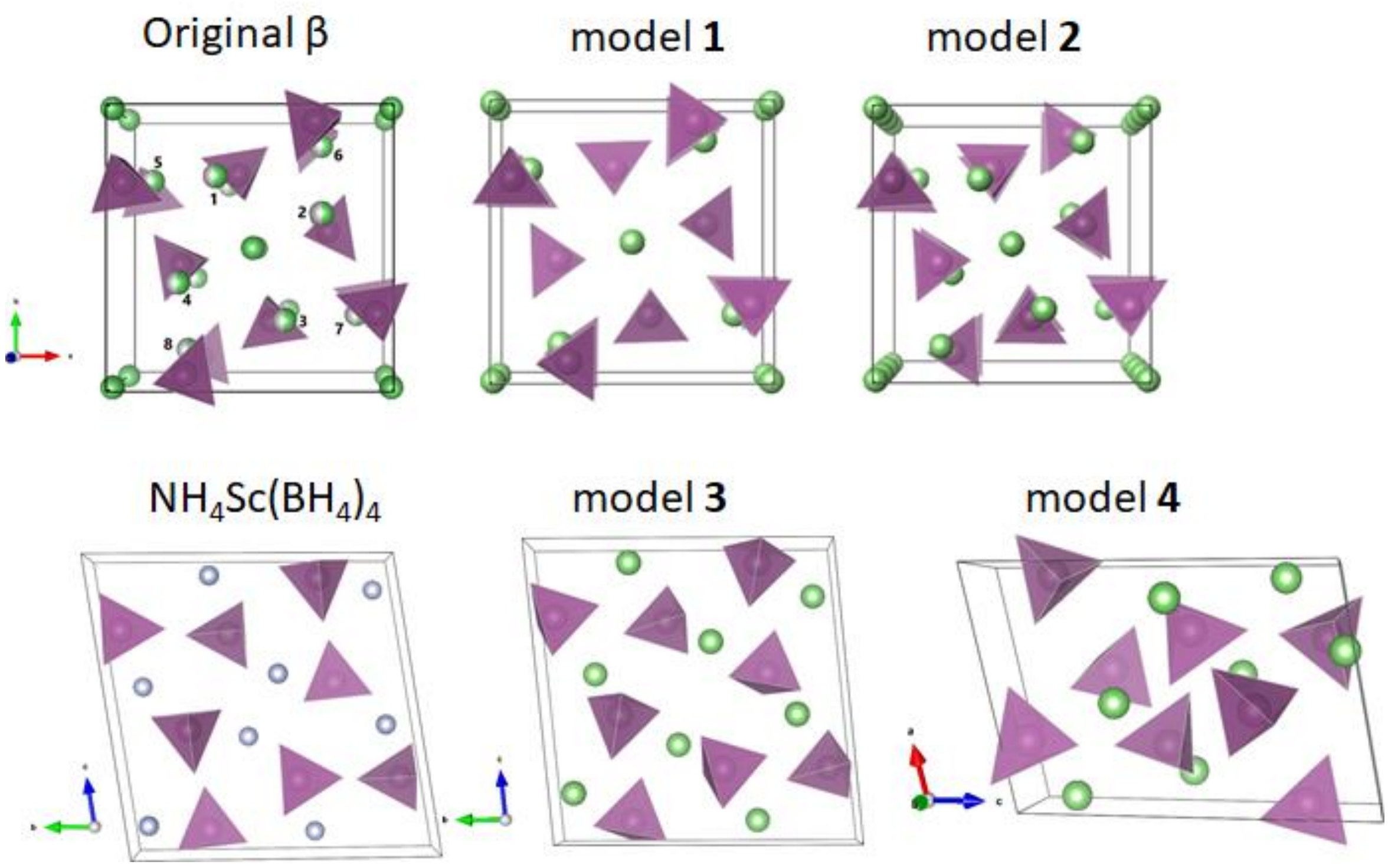

**Figure S5.** TOP - Unit cell of originally refined β phase (left) and its ordered models **1** (middle) and **2** (left) obtained by removing half of the partially ordered Li sites. BOTTOM – $NH_4Sc(BH4)_4$ unicell (left, viewed in representation of originally resolved β) and models **3** and **4** derived from it. Optimized on DFT-D3 level of theory. Color code: Li – (half)green balls, complex anions $[Sc(BH_4)_4]^-$ - purple tetrahedra, N – light blue balls. B/H atoms are omitted for clarity.

**Table S2**: DFT-D3 results, including crystallographic data and energies, for the optimized models of β-$LiSc(BH_4)_4$ compared with the originally resolved XRD parameters. Also, data for $NH_4Sc(BH4)_4$ are added for comparison (parameters in β representation).

| | **$NH_4Sc(BH_4)_4$ exp [3]** | **β exp** | **model 1 DFT-D3** | **model 2 DFT-D3** | **model 3 DFT-D3** | **model 4 DFT-D3** |
|---|---|---|---|---|---|---|
| **SPGR** | $P2_1/c$ | $I$ 4/m | $P$ 4/m | $P$ $4_2$/n | $P$ $2_1$/c | $P$-1 |
| **Z** | 8 | 8 | 8 | 16 | 4 | 8 |
| **a [Å]** | 15.683 | 14.284 | 13.476 | 13.475 | 15.081 | 11.304 |
| **b [Å]** | 15.683 | 14.284 | 13.476 | 13.475 | 15.081 | 11.467 |
| **c [Å]** | 7.886 | 7.376 | 8.256 | 16.700 | 6.785 | 15.543 |
| **α [°]** | 89.5 | 90 | 90 | 90 | 93.8 | 69.8 |
| **β [°]** | 90.5 | 90 | 90 | 90 | 86.2 | 88.8 |
| **γ [°]** | 81.4 | 90 | 90 | 90 | 84.8 | 62.5 |
| **V/Z [Å$^3$]** | 239.68 | 188.12 | 187.41 | 189.51 | 191.16 | 206.81 |
| **E/Z [eV]** | | --- | -96.26 | -96.34 | -96.55 | -96.32 |

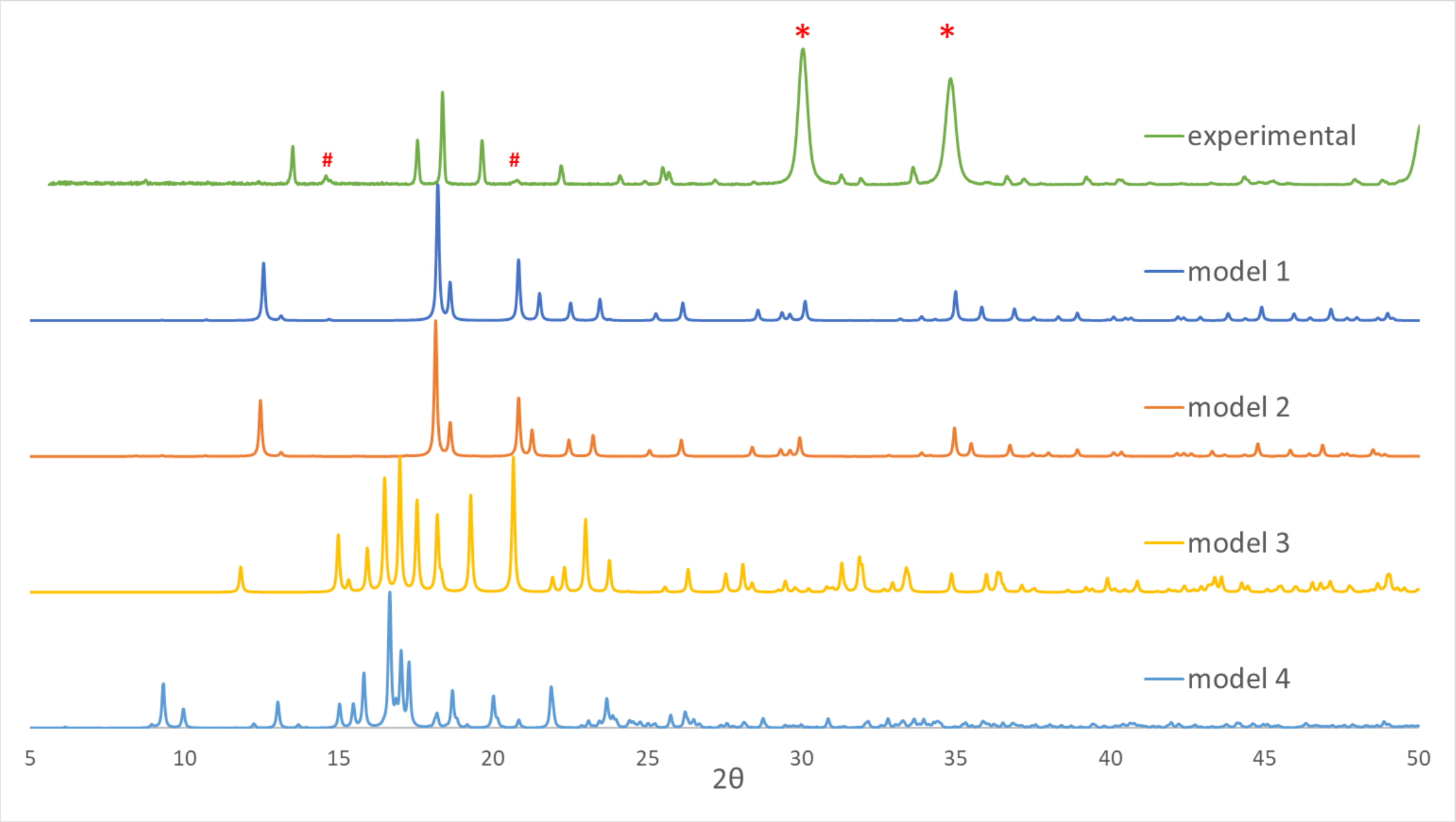

**Figure S6**: Comparison of the experimental and simulated XRD patterns for all β-$LiSc(BH_4)_4$ models optimized on DFT-D3 level. Red asterisks indicate LiCl reflections, while red hash signs indicate the most intense reflections of α-$LiSc(BH_4)_4$, both of them are present in the experimental pattern as by-products.

## S6. Redetermination of the beta phase: Rietveld fits of the diffraction patterns without and with partial $BH_4$ → Cl substitution

Having seen that none of the models proposed by theory may reasonably describe the beta phase, we have considered yet another possibility: that of partial $BH_4$ → Cl substitution. Below we show the fit assuming no substitution as well as the one where stoichiometry is a fitted parameter.

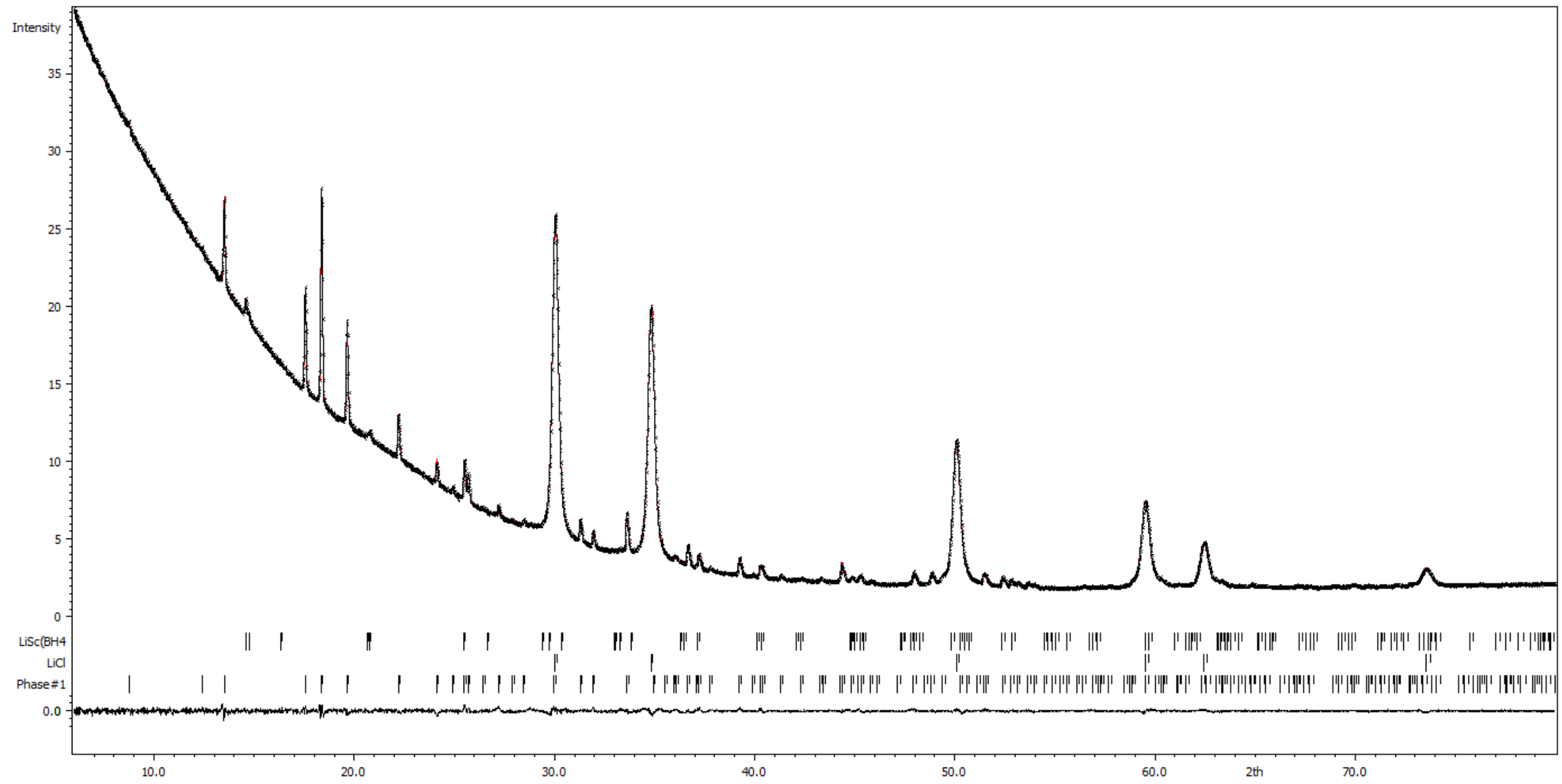

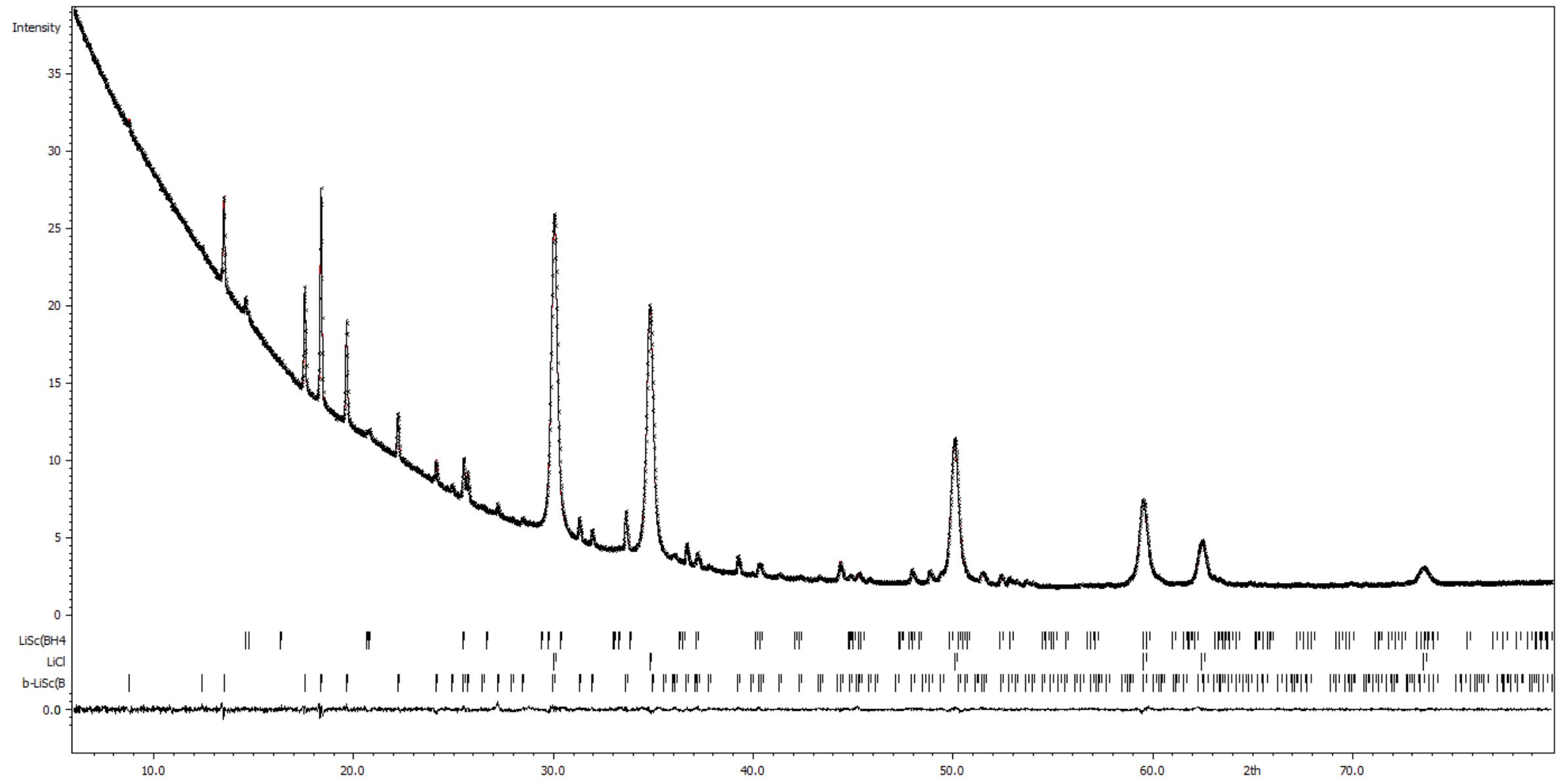

**Figure S7:** Rietveld plot of β-$LiSc(BH_4)_4$ at room temperature. The Bragg reflections of the crystalline phases are marked, from bottom to top: β-$Li[Sc(BH_4)_4]$, LiCl, α-$Li[Sc(BH_4)_4]$. Top plot – no $BH_4$-Cl substitution (wRp = 0.98%, cRp = 8.78%), bottom plot - $LiSc(BH_4)_{4-x}Cl_x$, $x \approx 0.7$ (wRp = 0.97%, cRp = 8.66%).

**Table S3.** Comparison of wRp/cRp of fits obtained for pure and Cl-substituted β phase.

| $LiSc(BH_4)_{4-x}Cl_x$ | x=0 | x≈0.7 |
|---|---|---|
| **wRp [%]** | 0.98 | 0.97 |
| **cRp [%]** | 8.78 | 8.66 |

## S7. The single crystal data for the disordered γ phase and its simulated powder diffraction pattern

**Table S4.** Crystallographic data for the disordered γ phase.

| | |
|---|---|
| **Composition** | $B_4 H_{16}$ Li Sc |
| ***M*/ g/mol** | 111.27 |
| ***T*/ K** | 100(2) |
| ***λ*/ Å** | 1.54184 (Cu $K_α$) |
| **Size[mm]** | 0.10 × 0.16 × 0.20 |
| **Crystal system** | cubic |
| **Space group** | *P*-43*m* |
| **unit cell parameters/ Å, °** | *a*=b=c=6.00500(10)<br>α=β=γ=90° |
| ***V* [$Å^3$]** | 216.540(11) |
| ***Z, $D_x$*/ g·$cm^{-3}$** | 1, 0.853 |
| ***μ* [$mm^{-1}$]** | 6.319 |
| ***F*(000)** | 60 |
| **$θ_{min}$, $θ_{max}$** | 7.349°, 70.358° |
| **Index ranges** | $-7 \leq h \leq 7$<br>$-7 \leq k \leq 7$<br>$-7 \leq l \leq 7$ |
| **Reflections collected/ independent** | 2325/ 111<br>[$R_{int}$=0.1298] |
| **Completness** | 100% |
| **$T_{max}$, $T_{min}$** | 0.400, 0.724 |
| **Data / restraints / parameters** | 111 / 1 / 10 |
| **GooF on $F^2$** | 1.245 |
| ***R* (all data)** | *R*1=0.0256<br>*wR*2=0.0640 |
| **$ρ_{max}$, $ρ_{min}$/ e·$Å^{-3}$** | 0.14, -0.18 |

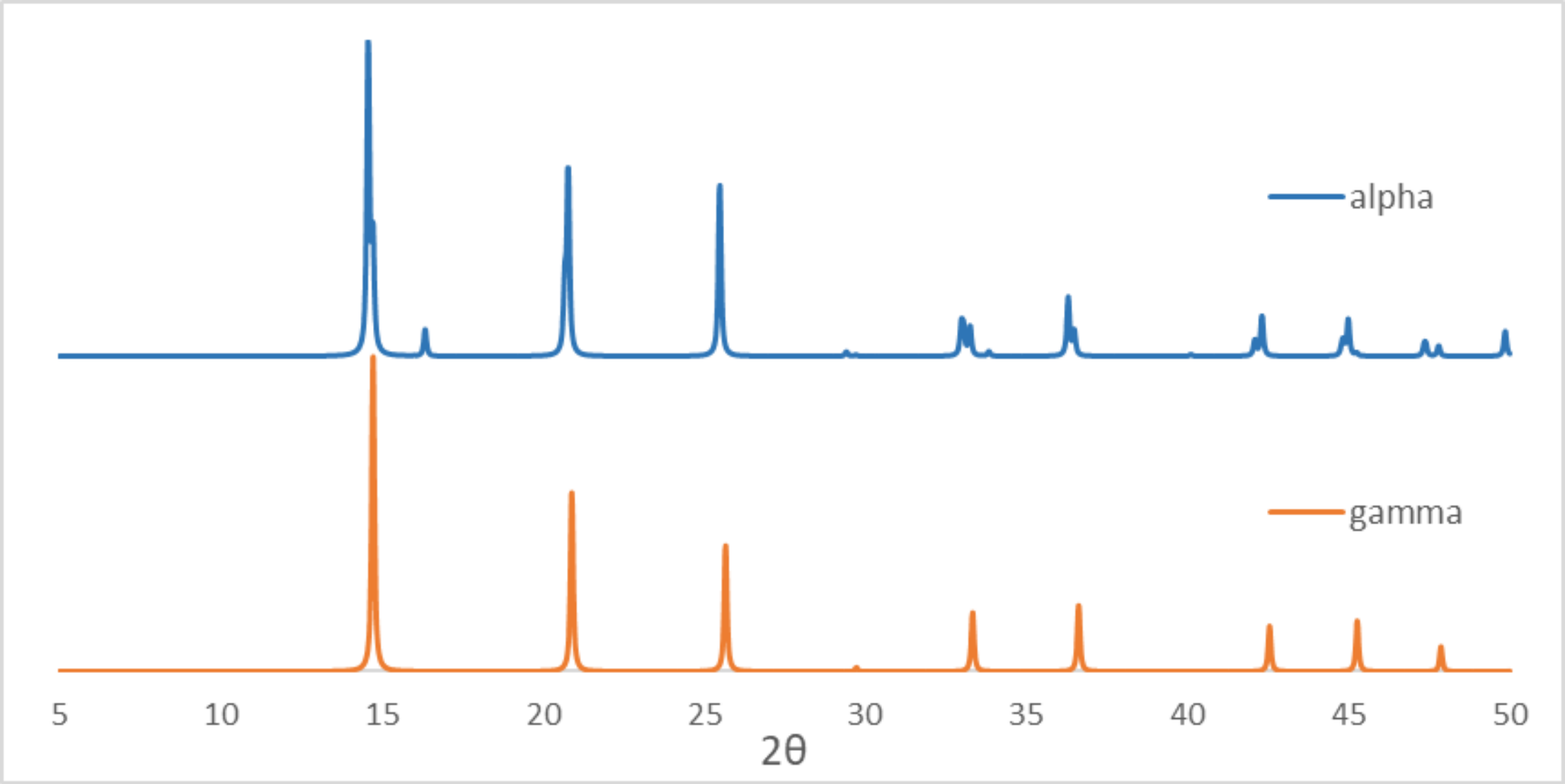

**Figure S8** Comparison of the simulated XRD patterns for ordered *P*-42c α model and disordered *P*-43m γ model.

## S8. CIF of computed ordered model of α

#=====================================================================

**Predicted ground state structure P-42c (ordered model of α)**

```
data_findsym-output
_audit_creation_method FINDSYM

_cell_length_a    5.8512000000
_cell_length_b    5.8512000000
_cell_length_c    11.8271100000
_cell_angle_alpha 90.0000000000
_cell_angle_beta  90.0000000000
_cell_angle_gamma 90.0000000000
_cell_volume      404.9193416304

_symmetry_space_group_name_H-M "P -4 2 c"
_symmetry_Int_Tables_number 112
_space_group.reference_setting '112:P -4 2c'
_space_group.transform_Pp_abc a,b,c;0,0,0

loop_
_space_group_symop_id
_space_group_symop_operation_xyz
1 x,y,z
2 x,-y,-z+1/2
3 -x,y,-z+1/2
4 -x,-y,z
5 y,x,z+1/2
6 y,-x,-z
7 -y,x,-z
8 -y,-x,z+1/2

loop_
_atom_site_label
_atom_site_type_symbol
_atom_site_symmetry_multiplicity
_atom_site_Wyckoff_label
_atom_site_fract_x
_atom_site_fract_y
_atom_site_fract_z
_atom_site_occupancy
_atom_site_fract_symmform
Li1 Li   2 d 0.00000 0.50000 0.25000  1.00000 0,0,0
Sc1 Sc   2 f 0.50000 0.50000 0.00000  1.00000 0,0,0
B1  B    8 n 0.68998 0.24408 0.88574  1.00000 Dx,Dy,Dz
H1  H    8 n 0.71273 0.19783 -0.01315 1.00000 Dx,Dy,Dz
H2  H    8 n 0.75391 0.44264 0.87050  1.00000 Dx,Dy,Dz
H3  H    8 n 0.75488 0.51924 0.36786  1.00000 Dx,Dy,Dz
H4  H    8 n 0.79344 0.11244 0.82647  1.00000 Dx,Dy,Dz
```

#=====================================================================

## S9. The secondary intermolecular H…H contacts in selected calculated models

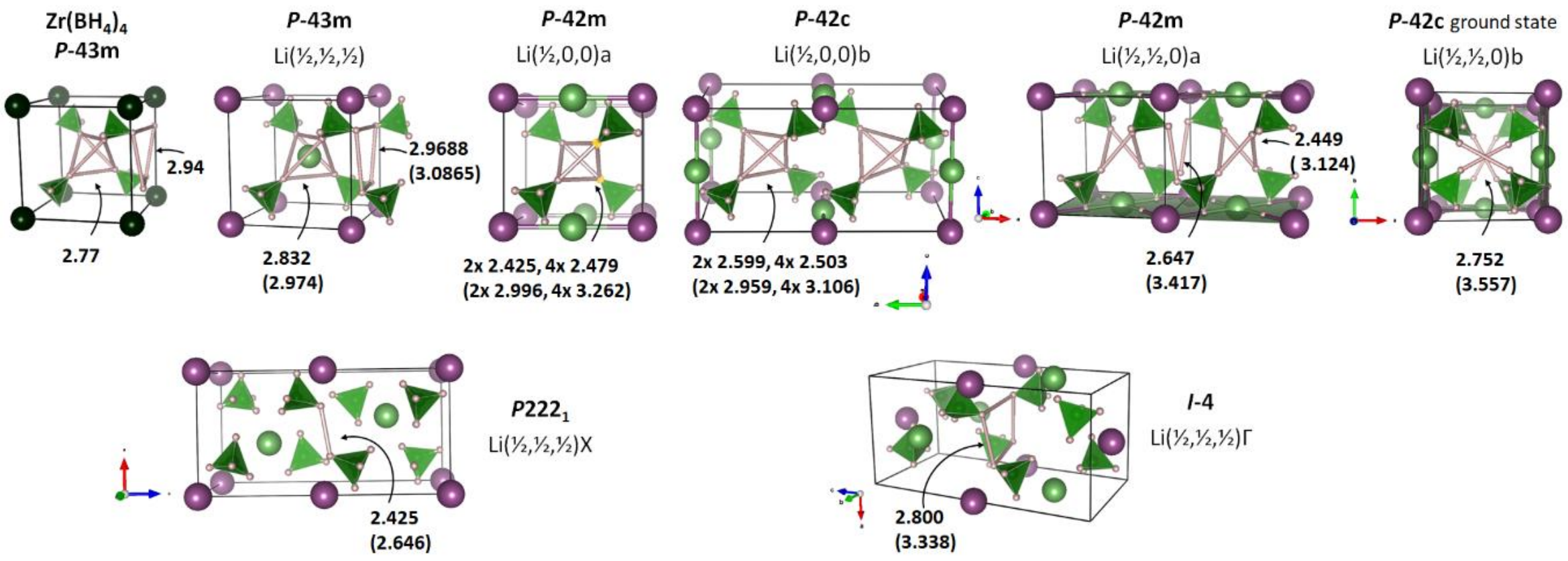

**Figure S9** The $Zr(BH_4)_4$ structure, the $Zr(BH_4)_4$ type models with Sc→Zr, Li→□ substitution and related phonon-mediated ones highlighting the shortest secondary intermolecular H…H contacts (light orange bonds) as calculated by DFT-D3 (DFT) method. Experimental values are provided for the $Zr(BH_4)_4$ structure [4]. Furthers details in the main text.